\documentclass{emulateapj}
\usepackage{times}
\usepackage{graphicx}

\slugcomment{Accepted to ApJ Letters}

\shorttitle{2FGL J2039.6$-$5620}
\shortauthors{Romani}

\begin{document}

\title{A binary counterpart for 2FGL J2039.6$-$5620}

\author{Roger W. Romani\altaffilmark{1}}
\altaffiltext{1}{Department of Physics, Stanford University, Stanford, CA 94305-4060,
 USA; rwr@astro.stanford.edu}


\begin{abstract}

We have identified an optical/X-ray binary with orbital period $P_b=5.47$\,h 
as the likely counterpart of the {\it Fermi} source 2FGL~J2039.6$-$5620.
GROND, SOAR and DES observations provide an accurate orbital period and 
allow us to compare with the light curve of an archival {\it XMM} exposure. Like many short-period
optical X-ray binaries associated with LAT sources this may be a interacting
(black widow/redback) millisecond pulsar binary. The X-ray light curve is consistent
with the emission associated with an intrabinary shock. The optical 
light curve shows evidence of companion heating, but has a peculiar asymmetric
double peak. The nature of this optical structure is not yet clear; additional optical
studies and, especially, detection of an orbital modulation in a $\gamma$-ray pulsar
are needed to elucidate the nature of this peculiar source.
\end{abstract}


\section{Introduction}

	As identification of the 1873 0.1--100\,GeV sources in the second {\it Fermi} 
Large Area Telescope (LAT) catalog \citep{2FGL} progresses, blazars and 
spin-powered pulsars increasingly dominate the associations \citep{3FGL}. For the brightest remaining 
unidentified sources gamma-ray spectral curvature and variability provide an excellent 
discrimination between these two possibilities \citep{r12}.  2FGL J2039.6$-$5620 = 3FGL 2039.6$-$5618
(hereafter J2039) is a relatively bright $F_{0.1-100\,{\rm GeV}}=1.7\pm 0.1\times
10^{-11}\,{\rm erg\,cm^{-2}\,s^{-1}}$ \citep[25.4$\sigma$,][]{3FGL} source with a good localization
at Galactic latitude -37$^\circ$,
a 3FGL ``variability index'' value of 34, and a ``curvature significance'' of 5.1$\sigma$.
Thus it is one of the least variable bright unidentified sources and displays significant
spectral curvature, placing it well within the pulsar zone of the curvature-variability plane.
Accordingly it has been repeatedly searched for radio \citep{cam15} and gamma-ray \citep{blind,pc14}
pulsations. The lack of any detection is best understood if the source is a millisecond
pulsar rapidly accelerated and/or wind-shrouded in a close, interacting binary 
\citep{rs11,r12,dmet14,roy15}.

	We have examined new and archival optical and X-ray data, discovering a short
period (5.47\,h) binary. The system shows evidence of heating, and the X-ray light curve
is reminiscent of other shrouded (black widow BW and redback RB) millisecond pulsar binaries.
The optical light curve is oddly double-peaked and asymmetric. Nevertheless we conclude
that this is another short period gamma-ray pulsar binary. With our well measured period and
position, this is a fruitful target for renewed pulse searches.

\section{Photometry and Orbital-Period Estimate}

	J2039 has a good 3FGL localization ($2.7^\prime$ radius) and we initially observed this field
as part of our LAT Unidentified source campaign, using the Goodman High Throughput Spectrograph (GHTS)
at the 4.2\,m SOAR telescope on 2013 August 10 (MJD 56514, UT dates are used throughout). A sequence
of $5\times(g^\prime r^\prime i^\prime)$ 180\,s images plus a 600\,s H$\alpha$ frame were obtained, 
spaced over 2 hours and largely covering the error ellipse. Despite relatively poor conditions, we were able
to search for variability for counterparts as faint as $g\sim 22$. One star (Figure 1), coincident with
an X-ray source in archival data (initially SWIFT, later XMM exposure, see below) was found to
vary in a correlated manner in all three bands, suggesting a characteristic 
period $\sim 3-6$\,h.  The USNO-B1.0 position of this source is 
$\alpha=20^{\rm h}39^{\rm m}34.987^{\rm s}$, $\delta=-56^\circ 17' 09.04''$ (2000.0, 
$\sigma_{RA}=0.06^{\prime\prime},\ \sigma_{DEC}=0.13^{\prime\prime}$). 
The catalog gives a proper motion estimate $\mu_{\rm RA} = 14\pm 4 {\rm mas\,y^{-1}}$ and
$\mu_{\rm RA} = -16\pm 9 {\rm mas\,y^{-1}}$. No H$\alpha$ structure was found in the field.

\begin{figure}[t!!]
\vskip 7.9truecm
\includegraphics{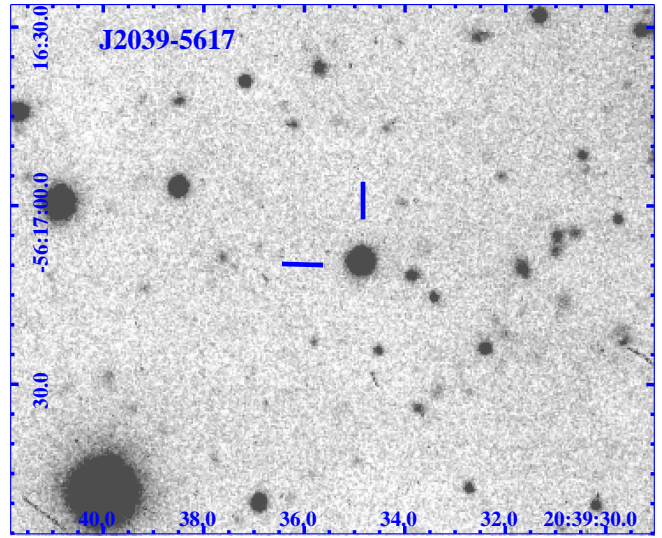}
\begin{center}
\caption{\label{Finder} 
Stacked SOAR/GHTS $g^\prime$ image of the J2039-5617 field.
}
\end{center}
\vskip -0.7truecm
\end{figure}

	We searched for archival observations in this direction and found an excellent
series of GROND \citep{get08} exposures of the J2039 error ellipse, with $72\times 115$\,s
$g^\prime r^\prime i^\prime z^\prime$(+JHK) exposures each on 3 days Aug 16-18 2014 
(MJD 56885-56887). Measuring
these optical frames, we obtain the light curves shown in figure 2. This data set 
showed a clear double-peaked light curve, with asymmetric maxima and minima and an orbital
period $P_B=0.22815\pm 0.00015$\,d (5.47\,h). The colors vary slightly with evidence for
heating (minimum $g^\prime-r^\prime$) at optical maximum.

	The quest for direct detection of blind pulsations is greatly helped by a precise
orbital period, which reduces the acceleration search space \citep{pet12}. Accordingly,
we compared with light curve phasing of our GHTS observations, 371\,d earlier. This
refined the period to 0.228116 $\pm N(1.4 \times 10^{-4})$\,d where the aliases
about the best minimum are set by orbit slips between the two observation sets.

	The target was covered in seven exposures in the DES early data release, 
in two frames each of $g^\prime$, $r^\prime$ and $i^\prime$, and a single $z^\prime$ frame.
We performed relative photometry against a set of the field stars from the GROND images, to
put the magnitudes on a consistent scale. The observation dates ranged from MJD 56538 to 
MJD 56558, between our other two epochs. Accordingly these points break the alias degeneracy,
clearly preferring $N=0$, the primary minimum. We thus establish $P_B=0.228116 \pm 0.000002$, with
barycentric epoch for optical maximum MJD 56885.085. As discussed below it is unclear how this
photometric epoch relates to the orbital kinematic epoch, i.e. $T_{ASC}$.

\begin{figure}[t!!]
\vskip 8.3truecm
\includegraphics{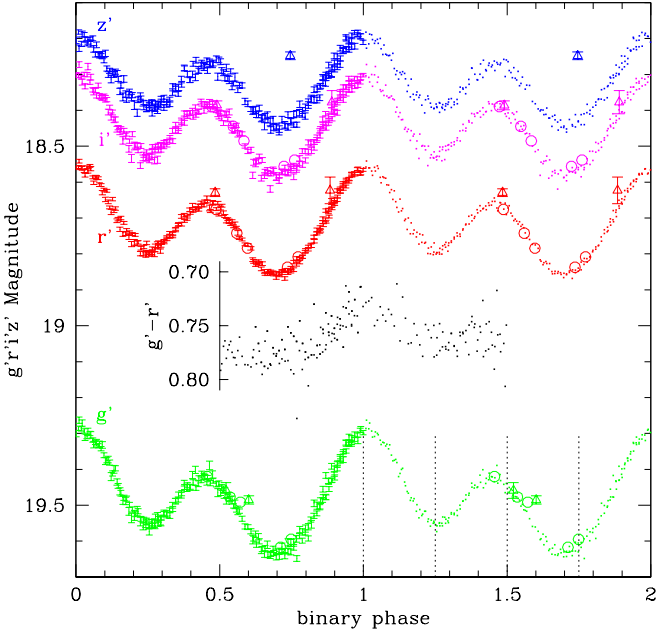}
\begin{center}
\caption{\label{LCfluxs} 
$g^\prime r^\prime i^\prime z^\prime$ light curves for J2039. Two cycles are shown with GROND
error bars plotted on the first cycle only. GHTS (circle) and DES (triangle points are also shown.
The DES $z^\prime$ offset may be due to the substantial differences in this filter.
Phase markers on the second period of the $g^\prime$ curve emphasize that the second maximum
and minimum are early by $\Delta \phi_B \sim 0.05$. A single period of the $g^\prime -r^\prime$
color is also shown, on a magnified scale.
}
\end{center}
\vskip -0.7truecm
\end{figure}

In Figure 2 the GHTS and DES points are seen,
calibrated against the GROND flux scale, showing good agreement with the mean light curve.
We also plot the DES $z^\prime$ point, which lies $-0.15$mag from the GROND $z^\prime$ curve.
This discrepancy may be cross-calibration error (GROND $z^\prime$ is split between DES $z^\prime$ and $Y$).
The DES data are however very spread in time: the $z^\prime$ point was 2.5 orbits off any 
other observation. The second $g^\prime$ point is also $\sim 3 \sigma$ from the expected 
flux and was obtained 5\,d before the first DES $g^\prime$ point. If, like many other BW/RB systems, the
companion undergoes occasional flares, this can explain these discrepancies. However the
small spread in GROND photometry over 3\,d and the otherwise good match of the GHTS and DES
points imply that the source is generally stable to better than $\sim 0.02$mag.

	J2039 is also covered in the Catalina Sky Survey (CSS;
http://www.lpl.arizona.edu/css/index.html; Drake et al. 2009) photometry archive with data extending
9.0\,y before the GROND exposures.  These unfiltered magnitudes with a typical error
$\sim 0.3$mag are too shallow to usefully probe the light curve modulation. However, they
do serve to confirm the source's general quiescence; out of 217 observations, only one point with
$\sigma_m <0.4$mag had $m<18$ (likely a measurement error in any event).

	The most remarkable aspect is of this light curve are the two asymmetric maxima and minima.
The first minimum follows the large maximum by $\Delta \phi_B\approx 0.25$, but the second maximum and minimum
arrive $\Delta \phi_B\approx 0.05$ too early to lie on the opposite side of the orbit (dotted lines
in Figure 2). The optical color
is bluest at optical maximum, but plateaus across the first minimum and second maximum,
decreasing to its reddest value at the second minimum. Clear there is off axis structure
in the system, with heating at two (nearly) opposite poles or, possibly an occultation at the
phase of the first minimum.

\begin{figure}[h!!]
\vskip 8.3truecm
\includegraphics{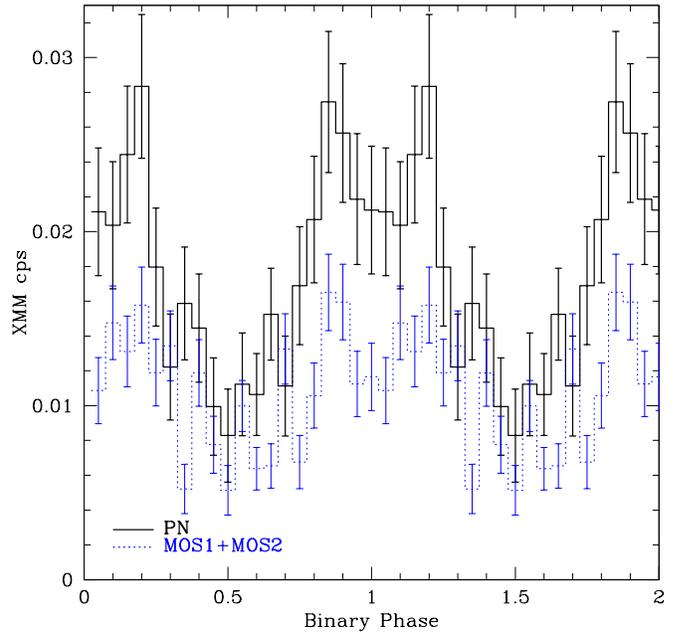}
\begin{center}
\caption{\label{X-rayLC} 
{\it XMM-Newton} light curves of J2039. Above PN, Below Combined MOS1+MOS2. $\phi_B=0$
is at optical maximum as in Figure 2.
}
\end{center}
\vskip -0.7truecm
\end{figure}

\section{Archival X-ray Light Curve}

	Archival SWIFT exposures show a few counts from this counterpart, but the
X-ray source is best measured in a 41.6\,ks live time {\it XMM-Newton} exposure (ObsID 0720650301,
PI: Mignani) on MJD 56575. This data set clearly shows the optical counterpart as the 
brightest source in the 3FGL error ellipse. With an observation between the GROND and GHTS epochs,
the predicted orbital phase is known to $\delta \phi_B=0.012$ dominated by the period uncertainty.
Folding the X-rays, we generated the PN and combined MOS1/2 light curves (figure 3). 
The peak is coincident with the optical maximum. The double structure, while of
modest statistical significance, is common to both instruments.

	The source is unresolved to {\it XMM} and delivered 0.024\,cps in 37\,ks 
effective PN camera exposure.  A power law spectral fit determines
a photon index $\Gamma=1.34_{-0.09}^{+0.10}$ and an upper limit on the absorption of 
$N_H < 1.9 \times 10^{20} {\rm cm^{-2}}$ (90\% errors). The corresponding unabsorbed
0.5-10\,keV flux is $8.2_{-0.5}^{+0.6} \times 10^{-14} {\rm erg\, cm^{-2} s^{-1}}$.
If a thermal component (e.g. from an underlying neutron star) is present a larger $N_H$
might be allowed. However present data do not require such a component and the absorbed
power law fit is quite adequate, with $\chi^2$/DoF=1.15.

\section{System Modeling and Conclusions}

	The source bears many similarities to several other short-period shrouded pulsar
binaries. At $P_B =5.47$\,h it may be either a redback or a black widow. However, at this
period it is much closer to the shrouding line in the former case and so it is more likely
a redback, given the lack of detection in repeated radio searches. Although the system properties 
will remain quite uncertain until a pulse detection is made, we can draw some tentative 
conclusions from the binary data.

	Our X-ray spectral fit indicates very little extinction, with the $N_H$ upper limit
corresponding to $A_V=0.07$ for the conversion of \citet{foiet15}. This is about half the total 
$A_V=0.16$ in this direction \citep{sf11}.
At this orbital period a main sequence companion would fill its Roche lobe for
$M_c = 0.61 M_\odot$ so we take this as an upper limit to  the companion mass. Such a star would
have $T_{eff} = 4180$\,K and $g^\prime-r^\prime=1.27$, so the observed color at minimum 
$g^\prime-r^\prime=0.78$ indicates appreciable heating. In fact, the colors are even bluer at maximum
with $g^\prime-r^\prime=0.73$ equivalent to $T_{eff}=4980$\,K. The origin of these increased
temperatures is unclear; the companion may be partly degenerate or it may experience tidal heating.
However in analogy with other shrouded MSP, it is likely that the effect of pulsar heating,
possibly mediated by an intrabinary shock, is dominant. If we consider just the temperature rise
to maximum, we derive an irradiation temperature $\sim 4200$\,K and for an orbital separation
$a=2.0(M_{tot}/2M_\odot )^{1/3}R_\odot$, an isotropic irradiation luminosity 
$4.2 \times 10^{33} {\rm erg\,s^{-1}}$. Lower companion masses require additional heating power,
although the relatively large $T_{Min}$ implies a low inclination angle or, possibly, 
rapid transport of heating flux to the companion night side, as invoked for some hot Jupiters
\citep[e.g.][]{hs15}.

	This heating flux is a lower limit to the true pulsar output. \citet{bet13} infer a typical
heating efficiency $\eta=0.1$ (some BW/RB appear to have larger heating efficiencies), which 
would imply a spin-down power ${\dot E} \approx 4 \times 10^{34} {\rm erg\,s^{-1}}$. 
Another power estimate can be derived from the gamma-ray flux $1.7 \times 10^{-11}
 {\rm erg\,cm^{-2}s^{-1}}$: \citet{psrcat} have noted a heuristic gamma-ray luminosity
$L_\gamma = (10^{33}{\rm erg\,s^{-1}} {\dot E})^{1/2}$, which gives us a distance estimate
$d \approx 1.25 {\dot E}_{34}^{1/4}$\,kpc. Note that the spindown power estimate for an
$\eta =0.1$ heating efficiency then places the source at 1.5\,kpc. Given that $N_H$ is below
the full Galactic column (at $b=-37^\circ$), we suspect a somewhat higher efficiency, lower
power and closer distance. The USNO B1.0 proper motion above corresponds to $v_\perp = 100 \pm 35 d_{kpc}
{\rm km\,s^{-1}}$. MSP binaries tend to be slower than the isolated young pulsar population,
with \citet{get11} giving $\langle v_\perp \rangle =90 {\rm km~s^{-1}}$. This
also supports a relatively close distance.

	The hard power-law X-ray spectrum and strong orbital modulation indicate a non-thermal
source. The light curve with an X-ray maximum flanked by two peaks and the observed 
$f_X/f_\gamma \approx 0.005$ are quite typical of BW/RB MSP showing evidence for
intrabinary shocks \citep{robet15}. 
These double peaks are generally centered around pulsar eclipse at superior conjunction
as for PSR J1959+2048 \citep{hkt12} or PSR J2239$-$0533 \citep{rs11}, but have also been 
seen at inferior conjunction e.g. PSR J2129$-$0429 \citep{robet15}.
The local minimum between these peaks may be a product of either companion eclipse or of beaming
along intrabinary shock surfaces. The low X-ray luminosity also suggests a binary in 
the pulsar-powered state \citep{lin14}.

	We close by discussing the most puzzling aspect of this system, the asymmetric double peaked 
optical light curve. Intriguingly, \citet{lht14} have recently measured the optical light curve of
the redback PSR J1628$-$3205, which has a 5\,h hour period and is near-Roche lobe filling. This
binary also shows asymmetric maxima and minima. These authors do not find a clear origin of this
asymmetry, but speculate on the possibilities of swept back intrabinary shocks or magnetically-directed
heating of spots on the companion surface. Both are plausible for J2049, as well. One difference, however,
is that \citet{lht14} place the optical maximum $\phi_B =0.5$ with respect to the pulsar ascending node.
Thus, with this phasing the star is brightest when viewed at {\it quadrature}. Accordingly
they assign the basic double-peak modulation to ellipsoidal variations, with the intrabinary shock
or magnetic heating supplying the required asymmetry. In our case without a pulsar ephemeris, we do not
know the true phases and for the moment arbitrarily set $\phi_B=0$ at the optical maximum. However 
as noted above the characteristic double peaked X-ray light curve is often along the pulsar-companion
axis. For J2049 this double peak is in phase with the optical maximum. If we interpret that maximum
as pulsar inferior conjunction (view of the heated face), then the double X-ray peak is similar to
that of PSR J2129$-$0429. Of course, none of these phase relationships are clear until we obtain
a kinematic ephemeris to reference our photometric epoch, either through a pulsar discovery
or through optical spectroscopy. We are pursuing both paths.
\bigskip
\bigskip
\bigskip

This research is based in part on observations obtained at the
Southern Astrophysical Research (SOAR) telescope, which is a joint
project of the Minist\'{e}rio da Ci\^{e}ncia, Tecnologia, e
Inova\c{c}\~{a}o (MCTI) da Rep\'{u}blica Federativa do Brasil, the
U.S. National Optical Astronomy Observatory (NOAO), the University of
North Carolina at Chapel Hill (UNC), and Michigan State University
(MSU).

\end{document}